\documentclass{phb-proc4-auth}

\usepackage{graphicx}
\usepackage{amssymb}

\begin{document}
\begin{frontmatter}


\journal{SCES '04}


\title{Quantum Phase Transitions in Spin-$\frac{1}{2}$ Ising Chain
       in Regularly Alternating Transverse Field:
       Spin Correlation Functions}

%
%
%
%
%
%

\author{Oleg Derzhko},
\author{Taras Krokhmalskii\corauthref{tk}}

%

\address{Institute for Condensed Matter Physics,
         National Academy of Sciences of Ukraine,
         1 Svientsitskii Str., L'viv-11, 79011, Ukraine}

%
%
%
%


%
%
%
%

\corauth[tk]{Corresponding Author:
            Institute for Condensed Matter Physics,
            National Academy of Sciences of Ukraine,
            1 Svientsitskii Str., L'viv-11, 79011, Ukraine.
            Phone: (0322) 76 19 78.
            Fax: (0322) 76 11 58.
            Email: krokhm@icmp.lviv.ua}


\begin{abstract}

We consider the spin-$\frac{1}{2}$ Ising chain in a regularly alternating transverse field
to examine the effects of regular alternation on the quantum phase transition
inherent in the quantum Ising chain.
The number of quantum phase transition points strongly depends
on the specific set of the Hamiltonian parameters
but never exceeds $2p$
where $p$ is the period of alternation.
Calculating the spin correlation functions numerically
(for long chains of up to 5400 sites)
and determining the critical exponents
we have demonstrated that two types of critical behavior are possible.
In most cases the square-lattice Ising model universality class occurs,
however, a weaker singularity may also take place.

\end{abstract}

%
%



\end{frontmatter}

%
%
%
%
%

We consider the spin-$\frac{1}{2}$ Ising chain in a transverse field
\cite{1,2,3}
assuming that the Hamiltonian parameters
(i.e. the nearest neighbor interactions and the on-site fields)
vary regularly along the chain
with a finite period of alternation $p$.
The Hamiltonian of the model reads
\begin{eqnarray}
\label {01}
H=\sum_n2I_ns_n^xs_{n+1}^x+\sum_n\Omega_n s_n^z
\end{eqnarray}
and the sequence of parameters in (\ref{01}) is
\begin{eqnarray}
I_1\Omega_1I_2\Omega_2\ldots I_p
\Omega_pI_1\Omega_1I_2\Omega_2\ldots I_p\Omega_p\ldots .
\nonumber
\end{eqnarray}
Recently,
it has been shown \cite{4}
that a regularly alternating transverse field
may strongly influence the quantum phase transition
inherent in the uniform chain \cite{1,2,3}.
In what follows we consider the chain of period 2
with $\Omega_{1,2}=\Omega\pm\Delta\Omega$ and $I_n=-1$.
Then the quantum phase transition points follow from the condition \cite{5}
\begin{eqnarray}
\label{02}
\left(\Omega^\star+\Delta\Omega\right)\left(\Omega^\star-\Delta\Omega\right)=\pm 1.
\end{eqnarray}
Eq. (\ref{02}) gives two critical fields
$\Omega^\star=\pm\sqrt{\Delta\Omega^2+1}$
if $\Delta\Omega<1$
or four critical fields
$\Omega^\star=\pm\sqrt{\Delta\Omega^2\pm 1}$
if $\Delta\Omega>1$.
Moreover,
if $\Delta\Omega=1$ in addition to $\Omega^\star=\pm\sqrt{2}$
we have one more critical field $\Omega^\star=0$.
Further we restrict ourselves to the case $\Delta\Omega=1$
and examine the critical behavior
in the vicinity of the two representative critical points
$\Omega^\star=\sqrt{2}$ and $\Omega^\star=0$.
This is by no means a restrictive example 
since only two types of critical behavior are expected for the model (\ref{01}) \cite{4}.
The critical behavior is characterized by a set of critical exponents.
Namely, the order parameter
(Ising magnetization) vanishes as $m^x\sim\left(\Omega^\star-\Omega\right)^\beta$.
The two-point spin correlation function
$\langle s_n^xs_{n+2r}^x\rangle-\langle s_n^x\rangle\langle s_{n+2r}^x\rangle$
in the limit $r\to\infty$ decays as
$\sim\left(2r\right)^{-\eta}\exp\left(-\frac{2r}{\xi}\right)$,
where $\xi\sim\vert\Omega-\Omega^\star\vert^{-\nu}$.
The transverse static susceptibility diverges as
$\chi^z\sim\vert\Omega-\Omega^\star\vert^{-\alpha}$
and the energy gap disappears as
$\Delta\sim\vert\Omega-\Omega^\star\vert^{\nu z}$.
From the analytical calculations \cite{4}
we know that in the vicinity of $\Omega^\star=\sqrt{2}$ we have
$\nu z=1$ and $\alpha=0$ ($\chi^z$ exhibits a logarithmic singularity)
whereas in the vicinity of $\Omega^\star=0$ we have
$\nu z=2$ and $\alpha=-2$
($\frac{\partial^2\chi^z}{\partial\Omega^2}$ exhibits a logarithmic singularity).
In the present paper we complete these analytical findings
by precise numeric results for other critical exponents $\eta$, $\beta$ and $\nu$
computing two-point spin correlation functions for chains of a few thousand sites \cite{6}.

\begin{figure}
\begin{center}
\includegraphics[width=0.780\linewidth]{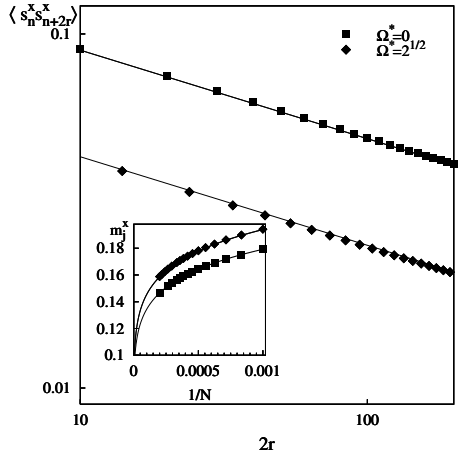}
\end{center}
\vspace{-1mm}
\caption{Power-law decay of spin correlations
at the critical points.
Main plot:
$\langle s_n^xs_{n+2r}^x\rangle$ vs $2r$
(log-log plot, $N=2000$, $n=920$,
$\Omega^\star=0$ (up),
$\Omega^\star=\sqrt{2}$ (down),
the solid lines are
$0.16\left(2r\right)^{-\frac{1}{4}}$
and
$0.08\left(2r\right)^{-\frac{1}{4}}$).
Inset:
$m_j^x$ vs $\frac{1}{N}$
($\Omega^\star=0$ (down),
$\Omega^\star=\sqrt{2}$ (up),
the solid curves are
$0.425N^{-\frac{1}{8}}$
and
$0.46N^{-\frac{1}{8}}$).
\label{f1}}
\end{figure}

We start our analysis
considering $\langle s_n^xs_{n+2r}^x\rangle$ at the critical fields
$\Omega^\star=0$ and $\Omega^\star=\sqrt{2}$.
The results reported in the main plot in Fig. \ref{f1}
clearly indicate the power-law decay of spin correlations
with $\eta=\frac{1}{4}$ at both critical points
$\Omega^\star=0$
(squares)
and $\Omega^\star=\sqrt{2}$
(diamonds).
We calculate the on-site $x$-magnetization $m_j^x$,
${m_j^x}^2=\lim_{r\to\infty}\langle s_j^xs_{j+2r}^x\rangle$,
for finite chains of $N$ sites
putting $j=\frac{1}{4}N$, $2r=\frac{1}{2}N$.
$m_j^x$ vanishes at both critical fields
as $N^{-\frac{1}{8}}$
as can be seen in the inset in Fig. \ref{f1}
(squares refer to $\Omega^\star=0$,
diamonds refer to $\Omega^\star=\sqrt{2}$).

\begin{figure}
\begin{center}
\includegraphics[width=0.980\linewidth]{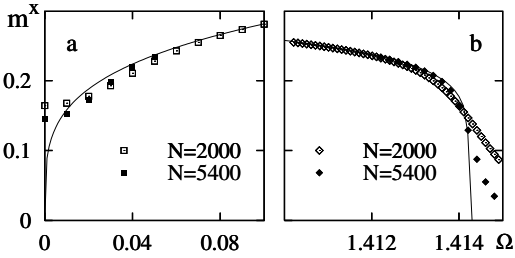}
\end{center}
\vspace{-1mm}
\caption{The order parameter behavior in the critical regions
($N=2000$: empty symbols, $N=5400$: full symbols)
\cite{4}.
The solid curves are
$0.5\Omega^\frac{1}{4}$ (a)
and
$0.5\left(\sqrt{2}-\Omega\right)^\frac{1}{8}$ (b).
\label{f2}}
\end{figure}

\begin{figure}
\begin{center}
\includegraphics[width=0.980\linewidth]{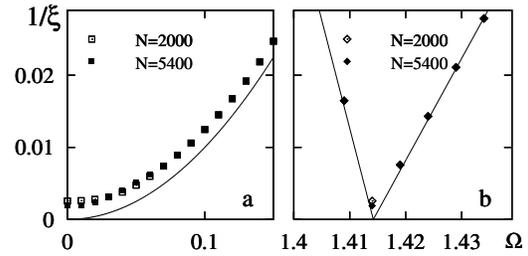}
\end{center}
\vspace{-1mm}
\caption{The inverse correlation length behavior in the critical regions
($N=2000$: empty symbols, $N=5400$: full symbols).
The solid curves are
$\Omega^2$ (a),
$3\vert\Omega-\sqrt{2}\vert$, $\Omega\le\sqrt{2}$
and
 $\sqrt{2}\vert\Omega-\sqrt{2}\vert$, $\Omega\ge\sqrt{2}$ (b).
\label{f3}}
\end{figure}

We turn to the critical region.
The results for $m^x=\frac{1}{2}\left(m_1^x+m_2^x\right)$
are shown in Fig. \ref{f2} (see also \cite{4}).
In the vicinity of $\Omega^\star=0$ (Fig. \ref{f2}a)
we observe a power-law vanishing of the order parameter with $\beta=\frac{1}{4}$.
In the vicinity of $\Omega^\star=\sqrt{2}$ (Fig. \ref{f2}b)
we observe a power-law vanishing of the order parameter with $\beta=\frac{1}{8}$.
Finally,
in Fig. \ref{f3}
we present the results for the correlation length.
While in the vicinity of $\Omega^\star=0$ we find $\nu=2$
(Fig. \ref{f3}a)
in the vicinity of $\Omega^\star=\sqrt{2}$ we clearly observe $\nu=1$
(Fig. \ref{f3}b).

To summarize, combining the analytical and numerical results we have
found that the spin-$\frac{1}{2}$ Ising chain in a regularly
alternating transverse field may exhibit the critical behavior which
belongs to the square-lattice Ising model universality class with
$\beta=\frac{1}{8}$, $\nu=1$, $\eta=\frac{1}{4}$, $\alpha=0$, $z=1$
(as the considered chain at $\Omega^\star=\sqrt{2}$) and a weaker
singularity with $\beta=\frac{1}{4}$, $\nu=2$, $\eta=\frac{1}{4}$,
$\alpha=-2$, $z=1$ (as the considered chain at $\Omega^\star=0$).
Let us finally note, that the spin correlation functions required
for the estimation of exponents $\beta$, $\nu$, $\eta$ should be
possible to calculate analytically as in the uniform case. We left
this problem for future studies.

\vspace{3mm}

The authors thank J.~Richter and O.~Zaburannyi
for collaboration in the earlier stage of this study
and Ja.~Ilnytskyi for discussions.
Preliminary results of this study were presented
at the 5th Small Triangle Meeting on theoretical physics
(Medzev, Slovak Republic, 2003).
O.~D. thanks the organizers for support
and M.~Ja\v{s}\v{c}ur and J.~Stre\v{c}ka for discussions.

%
%
%
%

%
%
%
%


\end{document}